\documentstyle[aps,epsf,epsfig]{revtex}
\def \be   {\begin{equation}}
\def \ee   {\end{equation}}
\def \l {\label}
\begin{document}
\input epsf
\baselineskip=25pt
\title{Discrete Fields on the lightcone}
\author{Manoelito M de Souza}
\address{Universidade Federal do Esp\'{\i}rito Santo - Departamento de
F\'{\i}sica\\29065.900 -Vit\'oria-ES-Brasil}
\date{\today}
\maketitle
\begin{abstract}
\noindent We introduce a classical field theory based on a concept of extended causality that mimics the causality of a point-particle Classical Mechanics by imposing constraints that are equivalent to a particle initial position and velocity. It results on a description of  discrete (point-like) interactions in terms of localized particle-like fields. We find the propagators of these particle-like fields and discuss their physical meaning, properties and consequences. They are conformally invariant, singularity-free, and describing a manifestly covariant $(1+1)$-dimensional dynamics in a $(3+1)$ spacetime. Remarkably this conformal symmetry remains even for the propagation of a massive field in four spacetime dimensions. The standard formalism with its distributed fields is retrieved in terms of spacetime average of the discrete fields. Singularities are the by-products of the averaging process. This new formalism enlighten the meaning and the problems of field theory, and may allow a softer transition to  a quantum theory.
\end{abstract}
\begin{center}
\end{center}

\section
{INTRODUCTION}

In a Minkowski spacetime of metric $\eta=diag(-1,1,1,1)$ the 4-vector separation $\Delta x(\tau)=x(\tau_{1})-x(\tau_{2})$   of two worldline events $x(\tau_{1})$ and $x(\tau_{2},)$ of a particle, parameterized by its propertime $\tau$, satisfies
\be
\label{A}
\Delta\tau^2=-\Delta x.\Delta x=-\Delta x^{2}.
\ee
In our notation an event x is given by its four-cartesian coordinates $({\vec x},t).$
 Besides being the definition of the particle propertime (\ref{A}) has also the role of describing causality.
For a massless particle $\Delta\tau=0,$ so its worldline must be parameterized by an affine parameter, and  
\be
\label{il}
\Delta x^{2}=0
\ee 
defines a local double (past and future) lightcone, $\Delta t=\pm|\Delta{\vec{x}}|.$ A massless physical object (particle or field) must remain on this three-dimensional-hypersurface lightcone. For a massive object $\Delta\tau\ne0,$ $\Delta t>|\Delta{\vec{x}}|$ and so it must remain inside its light cone. This is the well-known essence of the geometrical vision of the Einstein concept of local causality, which is a basic stone in the modern field theory building. It is essentially an access restriction to regions of the spacetime manifold: the lightcone and its interior form the allowed spacetime from each point.\\
An important aspect of this causality concept is that given an information of an event, for example, the position P of a given point physical object, all we are assured of, based only on (\ref{A}), is that this point object will remain inside (if it is massive) or on (if it is massless) the lightcone of vertex at P (the P-lightcone for short). This is the maximum that can be inferred as it was not given the object instantaneous velocity at P. It is nonetheless sufficient to attend the causality requirements of Quantum Mechanics and field theories as they are based on a concept of continuous fields diffused (or distributed) over the whole space-time or at least over the allowed space-time, i.e. the lightcone or its interior. The information that we had at P just diffuses in the allowed spacetime like a drop of ink in a glass of water.  In a field theory the time evolution of a system defines a Cauchy problem  which is the correspondent analogous to a classical-mechanics system whose evolution is determined by the initial positions and velocities of its elements. 
In point-particle classical mechanics the causality constraints come blended  with the very equation of motion; for a free point particle the Newton's law of inertia contains already all of its causality requirements. The complete information requires its initial position and velocity.

But if, for one side, the classical mechanics  narrowness cannot accommodate the modern physics data (particle's interference) that requires this continuous field concept even if having to quantize it later, for the other side, this concept of a continuous and distributed field faces well known difficulties on dealing with problems of self-interactions. The classical fields are not defined at their sources; their quantization is plagued with infinities that require mathematically troubled  renormalization processes, and even this fails for the gravitational field. It seems to us, that this approach of starting from a continuous and distributed field and trying then to get descriptions of discrete and localized object from it has already over-reached its stretched limits.\\
  In this paper we introduce a new  approach of defining a classical field theory based on discrete and localised fields, which are closer to the concept of a classical point particle. Hopefully this will avoid many of the problems that are intrinsic to a continuous and distributed field and will allow a softer transition to a quantum theory. This approach is based on a more strict concept of causality in classical field theory that is closer  to the causality concept  that exist in Classical Mechanics. We refer to it as EXTENDED CAUSALITY and it is introduced in Section II; its geometrical interpretation is briefed in the Appendix A.  The dynamical content of the causality constraints is discussed in section III.  We show, in Section IV, how the fields are defined in this formalism. The concept of a ``classical photon", i.e. a point element of a classical electromagnetic field,  is introduced as the point intersection of an electromagnetic wave front signal with a light cone generator. The Maxwell field now represents just the average effect of these ``classical photons"; its continuity is just a consequence of being an average and so are also its associated old problems like the charge infinite self-energy. In Section V we find (the actual calculation is transferred to the Appendix B) the propagator of a ``classical photon" whose main properties and physical consequences are discussed in Section VI and proved in the Appendices C and D. In Section VII we show how to retrieve the standard formalism and we conclude in Section VII with a qualitative discussion on the meaning and structure of a field theory born from this formalism.

\section{EXTENDED CAUSALITY}

We want to implement causality in field theory like in the way that it is done in Classical Mechanics, which requires informations on two neighboring events x and $x+dx$; that corresponds to giving the object position and its instantaneous velocity. This requirement, in terms of (\ref{A}), is equivalent to taking its differential
\be
\Delta\tau d\tau=-\Delta x.dx,
\ee
and taking it as a second causality restriction to be imposed simultaneously and on the same footing of the first one (\ref{A}).
Then, for $\Delta \tau\ne0$ (massive fields), we have that
\be
\label{dlcm}
d\tau+V.dx=0,\qquad\hbox{ for}\quad m\ne0,
\ee
with $V=\frac{\Delta x}{\Delta\tau}{\Big|}_{\tau_{s}},$ where $\tau_{s}$ is a solution of (\ref{A}) for a given initial condition $\tau_{0}$ and $x_{o}.$ For a massless field, $\Delta\tau=0$, and then (\ref{A}) and (\ref{dlcm}) are, respectively, replaced by (\ref{il}) and by
\be
\label{dlck}
d\tau+K.dx=0,\qquad\hbox{ for}\quad m=0,
\ee
where K is collinear to $\Delta x.$ Both K and $\Delta x$ are  null 4-vectors: $K^{2}=0;$ $(\Delta x)^{2}=0.$ 
For a massless field or in the limit of $d\tau\rightarrow0$ there is a clear geometric picture (the complete picture, valid for both $d\tau=0$ and $d\tau\ne0$, has been transferred to the Appendix A.) The constraint (\ref{dlck}) describes a hyperplane tangent to the light cone  (\ref{il}) and orthogonal to $k_{\mu}= -\frac{\partial\tau}{\partial x^{\mu}}$, where $\tau$, according to (\ref{A})< is seen as a given function of x. The intersection of (\ref{il}) and (\ref{dlck}) defines a lightcone generator orthogonal to $K_{\mu}$ and tangent to $K^{\mu},$ which is collinear to $\Delta x^{\mu}.$  $\;K^{\mu}k_{\mu}=0.$	 See Figure 1.\\

\vglue.5cm

\begin{minipage}[]{5.5cm}
\parbox[b]{5.5cm}
{
\begin{figure}
\epsfxsize=400pt
\hglue-2.0cm
\epsfbox{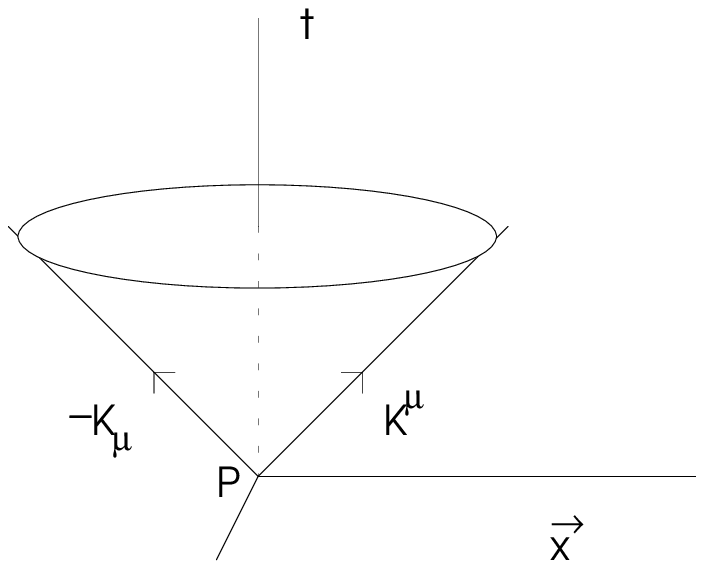}
\vglue-10cm
\end{figure}}
\end{minipage}\hfill
\hfill
\begin{minipage}[]{7.5cm}
\parbox[b]{7.5cm}{Fig. 1.The P-lightcone and its interior as the accessible spacetime for a physical object at P. $K^{\mu}=({\vec K},K^{4})$ and ${\bar K}^{\mu}=(-{\vec K},K^{4})$ are lightcone generators orthogonal to each other at the vertex. ${\bar K}^{\mu}=-K_{\mu}.$ $\; K_{\mu}=\eta_{\mu\nu} K^{\mu}.$}
\end{minipage}


\vglue-.5cm
For a generic situation we shall use 
\be
\l{dlc}
d\tau+f.dx=0,
\ee
in the place of (\ref{dlcm}) and (\ref{dlck}) where $f$is a constant 4-vector that stands for V if $f^{2}=-1$ or for  K if $f^{2}=0$. \\
 The intersection of (\ref{A}) and (\ref{dlc}),  with the elimination of $d\tau,$  corresponds to 
\be
\label{p}
(\eta_{\mu\nu}+f_{\mu}f_{\nu})dx^{\mu}dx^{\nu}=0.
\ee
The equation (\ref{p}) defines a projector $\Lambda_{\mu\nu}=\eta_{\mu\nu}+f_{\mu}f_{\nu},$ orthogonal to $f^{\mu},\;$ $\;(f^{\mu}=\eta^{\mu\nu}f_{\nu}),$ as $\Lambda_{\mu\nu}f^{\nu}\equiv0$ for  $f^{2}=-1,$ and $\Lambda_{\mu\nu}f^{\nu}=f_{\mu}$ for $f^{\mu}f_{\mu}=0.$ So, $\;f^{\mu}\Lambda_{\mu\nu}f^{\nu}=0,$ for all f. Therefore (\ref{p}) is a restriction that the displacements $dx$ be along a fiber tangent to $f^{\mu}.$ \\
The constraint (\ref{dlc}), with a choice for the origin, may be written as $\Delta\tau=-f_{\mu}x^{\mu}$ or  as 
\be
\l{dt}
-\Delta\tau=|{\vec f}|x_{\hbox{\tiny L}}+f_{4}t,
\ee
where $x_{\hbox{\tiny L}}$ is the coordinate along the ${\vec f}$-direction, $x_{\hbox{\tiny L}}=\frac{{\vec f}.{\vec x}}{|{\vec f}|}$. For fixed $\Delta\tau$ and $f,$ we have
\be
\l{dt1}
t=\frac{|{\vec f}|}{f^{4}}x_{\hbox{\tiny L}}+\frac{\Delta\tau}{f^{4}},
\ee
describing a hyperplane whose trace on the coordinate plane $(x_{L},t)$ is shown in Figure 2.
 

\parbox[]{7.5cm}{
\begin{figure}
\epsfxsize=400pt
\epsfbox{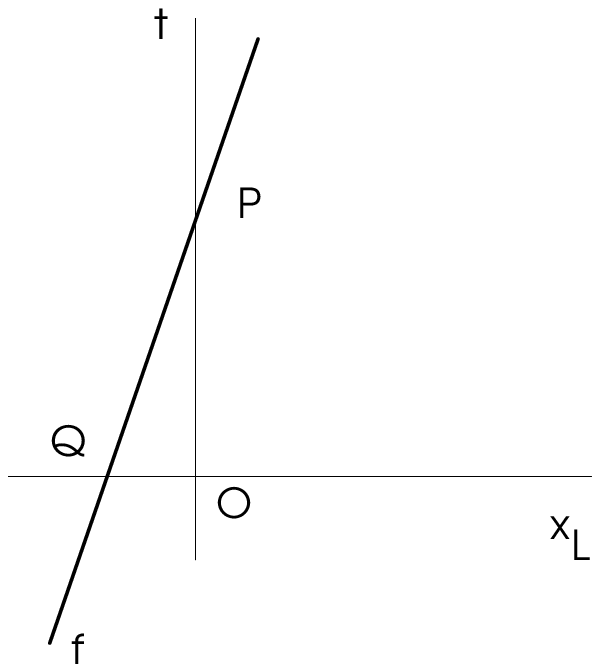}
\vglue-5cm
\end{figure}
\vglue-8cm}\\
\mbox{}
\hfill
\hspace{5.0cm}
\parbox[]{7.5cm}{\vglue-5cm Fig. 2.
The straight-line  support of PQ is the trace of the hyperplane described by $\Delta\tau+f.\Delta x=0.$ }\\ \mbox{}

Equation (\ref{A}) is the mathematical statement of the principle of local causality;  (\ref{A}) and (\ref{dlc}), together, describe this much more strict causality in field theory that we call EXTENDED CAUSALITY. Let us underline the differences between the local and the extended causality concepts. The imposition of (\ref{A}) defines, at each point P, a lightcone that divides the spacetime manifolds in two regions that, respectively, are and are not  causally connected to P, and in other words, restricts the allowed spacetime of a physical object at P to the P-lightcone or to its interior; this is the local or micro causality. The imposition of  (\ref{p}) defines, in this allowed spacetime a fiber, a straight line tangent to $f^{\mu}$, that represents the only left degree of freedom for a free physical object at P. This is the extended causality. \\ This will produce a field theory that mimics Classical Mechanics on its deterministic causality implementation.

\section{DYNAMICS AND CAUSALITY} 

The constraint (\ref{p}) defines a fiber, a unidimensional manifold along which a point particle moves or where a propagating point-like field is to be defined; it is their only left degree of freedom. For a constant four-vector $f$ this fibre is a straight line and so (\ref{p}), describing the propagation of a free point particle/field,  is just a kinematical constraint. The events $x$ and $x+dx$  in  (\ref{p}), may be any pair of events on a single worldline of a free physical object. This would imply that there is no special point on a worldline; all points would be equivalent. The constraint (\ref{p}) would be translationally invariant along the entire worldline. But the real world is not so uninteresting because if we do actually follow along the worldline of a physical object in either direction, or towards higher and higher or towards lower and lower values of the worldline parameter $\tau$, we must reach a final (an end or a beginning) point. In the real world there is no infinite  straight-lines associated to the propagation of physical objects. This final point must necessarily be an intersection with world-lines of other physical objects. Causality requires that there must be at least three world-lines on each intersection. This intersection always represents an interaction: the creation or the annihilation of a physical point object at a point on the fiber f. See the Figure 3, a pictorial representation of an interaction process in this formalism. 
These worldline intersections represent the only physical interaction among physical objects propagating along their respective world-lines. Neither worldline can  remain the same after such an intersection. Therefore, in the particular case when one of the two events in the definition of $dx$ is a worldline final point, the constraint (\ref{p}) acquires a dynamical content as it describes a fundamental interaction process. This intersection point is a singular point on the worldline because $f,$ its tangent four-vector, is not defined there either because the worldline starts there (a creation process), ends there ( a annihilation process) or just because it changes into a new fiber on a new direction (as in the emission or absorption of a photon by a charge, for example). See Figure 3. This singularity is in no way and no where associated to any infinity \cite{hep-th/9610028}.

\begin{minipage}[t]{5.0cm}
\parbox[]{5.0cm}{
\begin{figure}
\vglue-5cm

\epsfxsize=400pt
\epsfbox{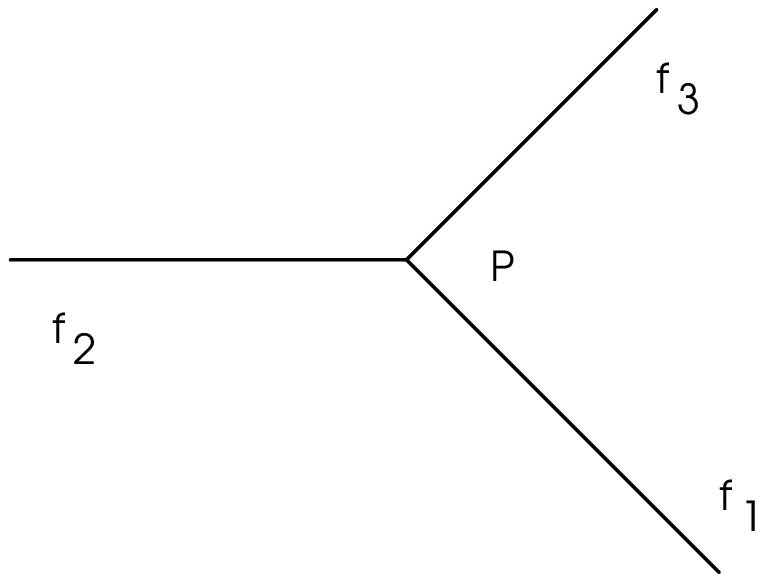}
\vglue-5cm
\end{figure}}
\end{minipage}
\hfill
\begin{minipage}[t]{5.0cm}
\parbox[]{5.0cm}{Fig. 3. 
Physical interactions are represented by the junction of , at least 3, world-lines at a point. Out of this isolated interaction point the world-lines represent free fields: they are straight lines as the $f'$ s are constant four-vectors.}
\end{minipage}

This interaction occurs at a single point, which is determined by the intersection of the world-lines. So it is a finite and extremely localised (point-like)  interaction.

\section
{FIELDS IN EXTENDED CAUSALITY}

We turn now to the question of how to describe a (3+1)-field in this (1+1)-formalism. Let us consider a spherically symmetric signal emitted at P, for fixing the idea. It propagates in/on the P-light-cone as a wave, distributed over an ever expanding three-sphere centred at P. It is a travelling spherical wave, an example of a distributed and non-localized field $A(x,\tau).$  Figure 4a is a spacetime diagram showing  three wave-fronts of a massless field A, represented by three circles (actually three spheres on a lightcone): $S_{1}$ at $t_{1}$, $ S_{2}$ at $t_{2}$ and $S_{3}$ at $t_{3}$. Figure 4b is the same representation in a three-space diagram.\\ 
\vglue-3cm

\parbox[]{5.0cm}{
\epsfxsize=400pt
\epsfbox{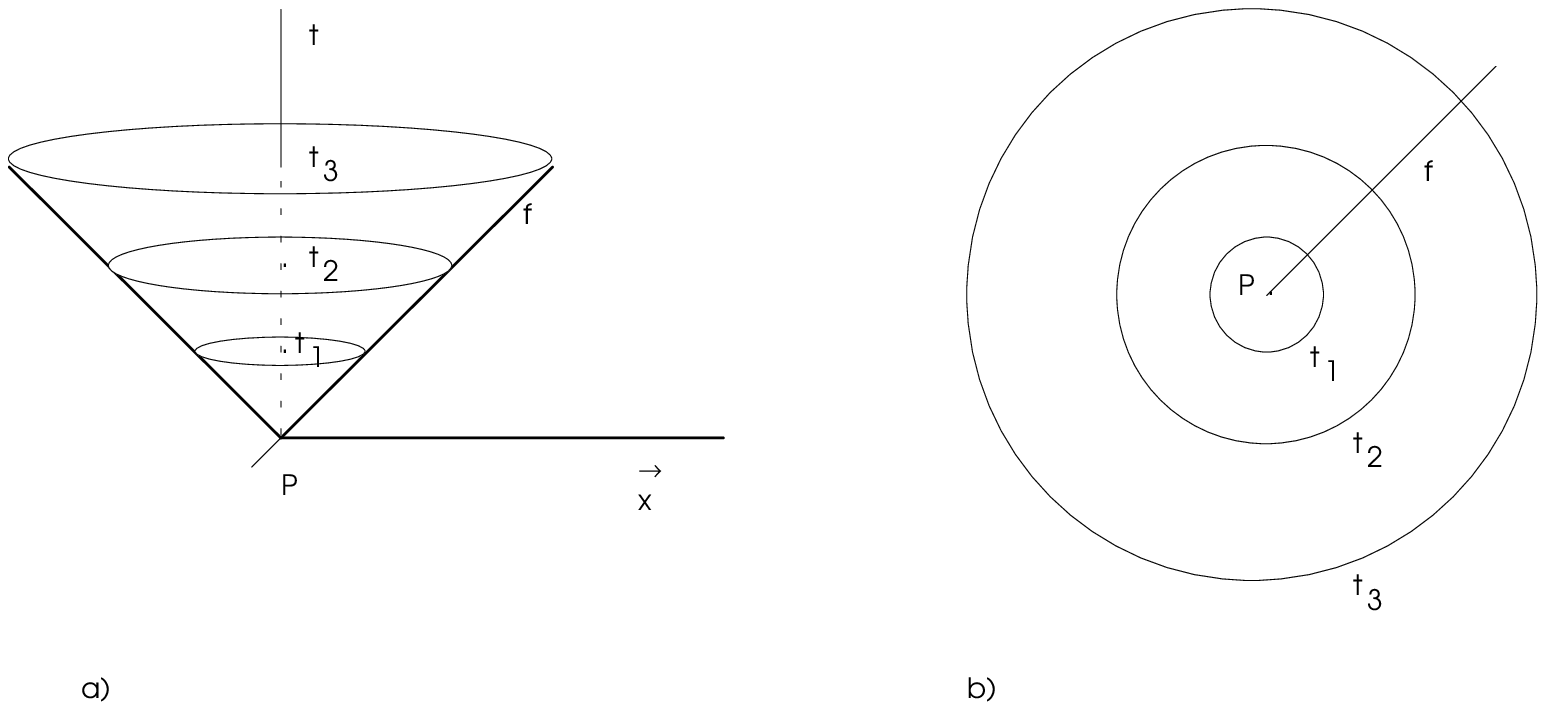}
\vglue-11cm}

\parbox[]{15.0cm}{Fig. 4. The front of a travelling spherical wave at three instants of time: (a) an spacetime diagram; (b) a three-space diagram. $f$ is a cone generator.}
\vglue1cm

 For the implementation of extended causality it is necessary to be supplied at each point P with a generic direction in spacetime, which we label by a four-vector $f$, and its $f$ fiber, that is the straight-line that passes through P and is tangent to f. This fiber is represented in Figures 4a and 4b by the straight line f. Let $A_{f}$ be the intersection of the wave front of A  with the fiber f. In other words,
\be
\label{af}
A(x,\tau)_{f}=A(x,\tau){\Big |}_{d\tau+f.dx=0}.
\ee
The symbol $A_{f}$ or $A(x,\tau)_{f}$ denotes the restriction of $A(x,\tau)$ to the fiber f. It represents an element of $A(x,\tau)$, the part of $A(x,\tau)$ contained in the fiber f: a point propagating along f. On the other hand, $A(x,\tau)$ represents the collection of all such elements $A(x,\tau)_{f}$ from all possible fibres $f,$ and so the converse of (\ref{af}) is given by
\be
\label{s}
A(x,\tau)=\frac{1}{4\pi}\int d^{2}\Omega_{f}A(x.\tau)_{f},
\ee
where the integral represents the sum over all directions of $f$ in or on  a lightcone, according to $f^{2}=-1$ or $f^{2}=0,$ respectively. $4\pi$ is a normalization factor. $A(x,\tau)$ is a continuous field and so $A(x,\tau)_{f}$ could also be seen as a continuous function of its parameter $f,$ but we prefer to draw another more interesting picture. We know from modern physics that this continuity of A is just an approximation. An electromagnetic wave, for example, is made of a large number of photons and, in a classical description, it can be seen as a bunch of massless point particles swarming out from P at all directions. We want to associate $A(x,\tau)_{f}$ to the point particle emitted at P in the direction $f$ and we will call it the CLASSICAL QUANTUM (as contradictory as it may appear to be)  at the fiber f. If A denotes an electromagnetic wave, $A_{f}$ will denote a CLASSICAL PHOTON. 

This represents a drastic change in the meaning of $A(x,\tau)$ and a reversal of what is taken as the primitive and the derived concept.  $A(x,\tau)$, while taken as the primitive concept, represented the actual physical field; now it has been  reduced to the average effect of a large, but finite, number of (classical) photons. $A(x,\tau)_{f}$ is a classical representation of the actual physical agent of electromagnetic interactions, the photons. $A(x,\tau)$ being, in this new context, just an average field representation of the exchanged photons can  produce good physical descriptions only at the measure of a large number of emitted and randomly distributed photons. This is the new interpretation of equation (\ref{s}). It does not make much difference for most of the practical situations that correspond, for example, to the description of an electromagnetic field involving a large number of photons and in a point far away from its sources.\\
It certainly fails for very low intensity light involving few photons. Imagine an extreme case of just one photon emitted, pictured as the fiber $f$ in the Figure 5.
Then $A(x,\tau),$  represented by the dotted circle, gives a false picture of isotropy while the true physical action (the photon) goes only along the $f$ direction. This situation is unavoidable when we approach to the field source, the electric charge, to such a small distance that the time elapsed between 
a photon emission and its detection is of the order of the period of time between two consecutive emissions of photons by the accelerated charge. In this extreme situation the description in terms of the averaged field $A(x,\tau)$, i.e. the usual field of electrodynamics , fails. The Gauss's equation has only meaning for a time-and-space averaged electric field. This is the origin  of all problem (infinite self-energy, the Lorentz-Dirac as an equation of motion, etc) for the usual local-causality-field theory  \cite{Rohrlich,Jackson,Parrot,Teitelboim,Rowe,Lozada}, and has already been discussed in \cite{hep-th/9610028}. See also \cite{hep-th/9610145} for an anticipated qualitative discussion of the meaning of field singularity and of static field in this context.\\

\begin{minipage}[]{5.0cm}
\parbox[b]{5.0cm}{
\begin{figure}
\vglue-6.5cm
\epsfxsize=400pt
\epsfbox{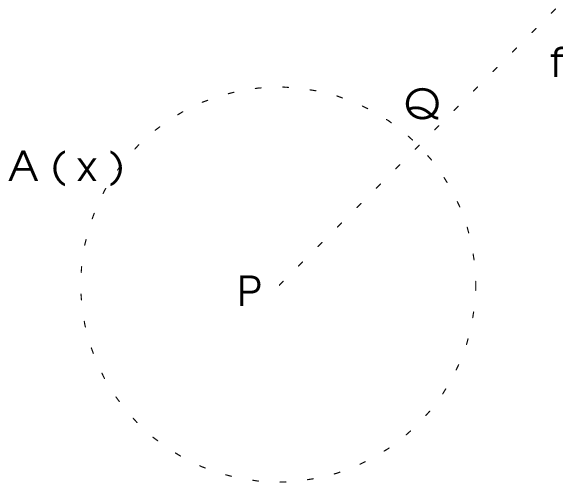}
\end{figure}}
\end{minipage}\hfill
\hfill
\vglue-12cm
\begin{minipage}[]{7.0cm}\hglue7.0cm
\parbox[b]{7.0cm}{Fig. 5. A very low intensity light with just one photon. The dotted circle represents the Maxwell field for this light. It transmit a false idea of isotropy. The point Q on the straight line $f$ represents the classical photon $A_{f}$.}\\ \mbox{}
\end{minipage}
\vglue2cm

\section
{FIELD EQUATION AND GREEN'S FUNCTION}

What we are proposing here is the development of a formalism of field theory where $A(x,\tau)$ is replaced by $A_{f}(x,\tau)$ in its role of a basic field, that is, to replace the usual local-causality field formalism by another one with the explicit implementation of (\ref{p}) instead of just  (\ref{A}). A characteristic of this new formalism is that the fields must be explicit functions of x and of $\tau,$ where $\tau$, as a consequence of the causality constraint (\ref{p}), is a supposedly known function of x, a solution of (\ref{A}). For a massless field, as it propagates without a change on its proper time, $\Delta\tau=0$, $\quad\tau$ is actually  the instantaneous proper-time of its source at the event of its emission. See the Figure 6 where $z(\tau)$ is the source worldline parameterized by its proper time $\tau.$ It pictures $A_{f}(x,\tau_{s}),$ emitted at $z(\tau_{s})$ and propagating along the fiber $f$ to the point x where it is observed (or detected).

\parbox[t]{5.0cm}{\vglue-5cm
\begin{figure}
\vglue-2cm
\epsfxsize=400pt
\epsfbox{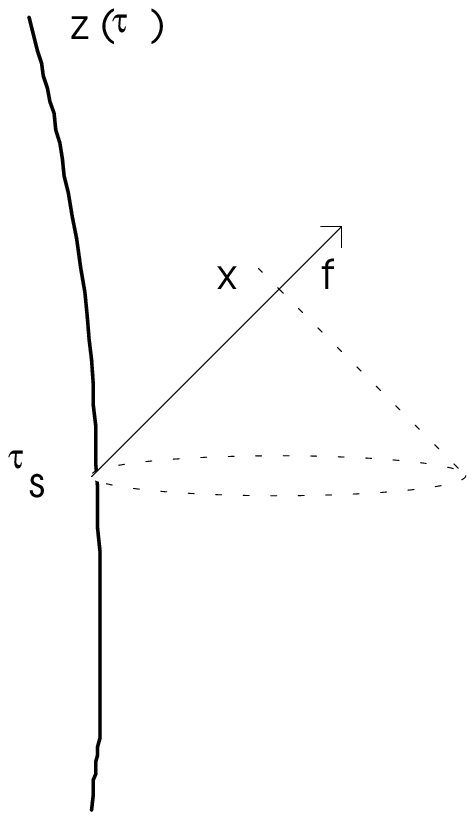}
\vglue-4cm
\end{figure}}\\ 
\mbox{}
\hspace{8.0cm}
\parbox[t]{7.0cm}{\vglue-7cm \hspace{5.0cm}\parbox[t]{7.0cm}{
Fig.6. A classical photon $A_{f}(x,\tau_{s})$
 emitted at $\tau_{s}$ propagates along a $f$ 
lightcone generator towards the point $x$.}}\\ \mbox{}
\vglue-3cm

The behaviour of a massless field $A_{f}(x,\tau_{s})$ is determined by its wave equation
\be
\label{we}
\eta^{\mu\nu}\frac{\partial^{2}}{\partial x^{\mu}\partial x^{\nu}}A_{f}(x,\tau)=J(x,\tau),
\ee
where $J(x,\tau)$ is the four-current source of $A(x,\tau)$.  But the constraint (\ref{p}) on $A(x,\tau)$ implies that
\be
\label{fd}
\frac{\partial}{\partial x^{\mu}}A_{f}=(\frac{\partial }{\partial x^{\mu}}+\frac{\partial \tau}{\partial x^{\mu}}\frac{\partial }{\partial \tau})A(x,\tau){\Big |}_{d\tau+f.dx=0}=(\frac{\partial }{\partial x^{\mu}}-f_{\mu}\frac{\partial}{\partial \tau})A(x,\tau){\Big |}_{d\tau+f.dx=0},
\ee
where we made use of (\ref{dlc}) for $f_{\mu}=-\frac{\partial\tau}{\partial x^{\mu}}{\Big |}_{d\tau+f.dx=0}.$
It is convenient to work with 5 independent variables, assuming that $\tau$ is independent of x, but replacing the ordinary derivative $\partial_{\mu}$ 
by  the operator $\nabla_{\mu}=\partial_{\mu}-f_{\mu}\partial_{\tau}.$
$$\partial_{\mu}\Rightarrow\nabla_{\mu}$$
$$\partial_{\mu}A(x,\tau){\Big |}_{d\tau+f.dx=0}\Rightarrow\nabla_{\mu}A_{f}(x,\tau)$$
The meaning of $\nabla_{\mu}$ is clear: it is the generator of the displacements allowed by the constraint (\ref{p}). It is a derivative along the fiber f. The $f$ in the definition of $\nabla$ is provided by the constraint (\ref{p}) over the field on which it acts. So, for example,
\be
\nabla_{\mu}(A_{f}B_{f'})=B_{f'}(\partial_{\mu}-f_{\mu}\partial_{\tau})A_{f}+A_{f}(\partial_{\mu}-f'_{\mu}\partial_{\tau})B_{f'}.
\ee
The field equation (\ref{we}) now re-written as
\be
\label{wef}
\eta^{\mu\nu}\nabla_{\mu}\nabla_{\nu}A(x,\tau)_{f}=J(x,\tau),
\ee
has solutions that we may write in terms of a Green's function as
\be
\label{sgf}
A_{f}(x,\tau_{x})=\int d^{4}yd\tau_{y}\; G_{f}(x-y,\tau_{x}-\tau_{y})\;J(y),
\ee
with $G_{f}(x-y,\tau_{x}-\tau_{y})$ being a solution of
\be
\label{gfe}
\eta^{\mu\nu}\nabla_{\mu}\nabla_{\nu}G_{f}(x-y,\tau_{x}-\tau_{y})=\delta^{4}(x-y)\delta(\tau_{x}-\tau_{y})\equiv\delta^{5}(x-y),
\ee
with
\be
\label{delta5}
\delta^{5}(x)=\frac{1}{(2\pi)^{5}}\int d^{5}p\; e^{ip_{M}x^{M}},
\ee
for $M=1,2,3,4,5$ and with $x^{4}=t$ and $x^{5}=\tau_{x}.$\\
In order to keep the physical discussion transparent, we have put the calculation into the Appendix B and simply quote the result here:
\be
\l{pr9'}
G( x,\tau_{x})_{f}=\frac{1}{2}\theta(b{\bar f}. x)\theta(b\tau)\delta(\tau+f. x),
\ee
or equivalently by 
\be
\label{pr9}
G_{f}(x,\tau)=\frac{1}{2}\theta(at)\theta(b\tau)\delta(\tau+f.x);
\ee
where $a,b =\pm1,$ and are constrained by
\be
\l{abe'} 
abf_{4}=|f_{4}|.
\ee 
f and ${\bar f}$ are defined by $f^{\mu}=({\vec f},f^{4})$ and ${\bar f}^{\mu}=(-{\vec f},f^{4}),\;$ ${\bar f}^{\mu}=-\eta_{\mu\nu}f^{\nu}.$  The Heaviside function $\theta(x)$ is defined by

\be
\theta(x)=\cases{
1,&if $x>0$;\cr
1/2 & if $x=0$;\cr
0   &  if $x<0$\cr}.
\ee

 \section{PROPERTIES OF $G_{f}$.}

Let us list here some of the properties of $G_{f}$.

\begin{itemize}
\item Although we have a $(3+1)D$ spacetime the presence of $\delta(\tau+f.x)$ in (\ref{pr9}) shows that the $A_{f}$ dynamics and kinematics are effectively $(1+1)D$; only the space direction along the ${\vec f}$ direction is relevant. The transversal coordinates do not participate; they are kept unchanged. So, $A_{f}$ is a point perturbation that propagates along the fiber f; it is not a distributed field like $A(x,\tau)$ that is defined and propagates on the entire lightcone.
\item Another remarkable property of (\ref{pr9}) is the absence of singularity; its delta-function argument is linear on the coordinate, in contrast to the one in the Liènard-Wiechert propagator ($G(x)=\theta(\varepsilon t)\delta(x^{2})$) which is proportional to $x^{2}$, and so has a singularity at the space coordinate origin. So, the great difference, with respect to singularity, is that the worldline of $A_{f}$ has end-points (points of creation or annihilation), see the Figure 3, but no singular points in the sense of infinities; the end-point is singular only in the sense of not having a well defined tangent. No infinity is ever involved \cite{hep-th/9610028}.
Geometrically it means that the support manifold of $A_{f}$ is a straight line, a complete manifold without singularity, while for $A(x,\tau)$ the support manifold is the lightcone, which is not a complete manifold because of the singularity at its vertex.
\item The equation 
(\ref{A}) corresponds to $t^{2}=\tau^{2}+ ({\vec x})^{2}$ and so $|t|\ge |{\vec x}|$
Then, with (\ref{abe'}) we have that $$\theta(b{\bar f}.x)=\theta[b(f_{4} t-{\vec f}.{\vec x})]=\theta[bf_{4}(t-\epsilon|{\vec x}|)]=\theta(bf_{4} t)=\theta(a t).$$ This justifies (\ref{pr9}). On the other hand, using the delta-function properties in (\ref{pr9'}) we can replace $\theta(b\tau)$ by $\theta(-bf.x)=\theta(-bf_{4}t)$. So the two theta-functions in (\ref{pr9}) require that
\be
at\ge0
\ee
and
\be
-bf_{4}t\ge0,
\ee
and their product implies on (\ref{abe'}), as $a,b=\pm1.$
\item For a fixed ${\vec x}$ there are  two associated propagator solutions, one for $\Delta t>0$ and another one for $\Delta t<0$, or equivalently $b=+1$ and $b=-1$. 
They are depicted in the Figure 7.
\hspace{-3cm}
\vglue-3cm

\hspace{-1cm}
\parbox[t]{5.0cm}{
\begin{figure}
\vglue-3cm
\epsfxsize=400pt
\epsfbox{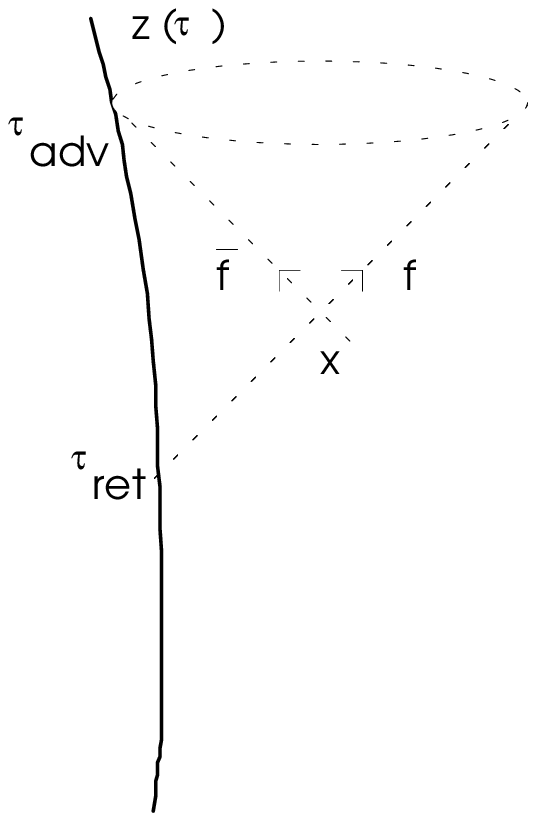}
\vglue-5cm
\end{figure}}\hfill
\\ \mbox{}
\hspace{5cm}
\parbox[t]{8.0cm}{\vglue-5cm 
Fig. 7. The The Li\`enard-Wiechert solutions as creation and annihilation of particles. There are two classical photons at the point x: one created at $\tau_{ret}$ and has propagated to x on the lightcone generator f; the other one propagating on a lightcone generator ${\bar f}$ from x towards the electron worldline where it will be annihilated at $\tau_{adv}$.}\\ \mbox{}
\vglue-2cm

They represent, respectively, the creation and the annihilation of a classical photon by the electric charge at its worldline $z(\tau).$ At the event x there are
two classical photons. One, that was emitted by the electron current $J$, at
$z(\tau_{ret})$ with $x^{4}>z^{4}(\tau_{ret})$, and is moving in the f
generator of the x-lightcone, $f^{\mu}=({\vec f},f^{4}).$  $J$ is its
source. The other one, moving on a ${\bar f}$-generator, ${\bar
f}^{\mu}=(-{\bar f},f^{4}),$ will be absorbed by $J$ at $z(\tau_{adv}),$ with
$x^{4}<z^{4}(\tau_{adv}).$ $J$ is its sink. See the Figure 7. They are both
retarded solutions and correspond, respectively, to the creation and
destruction of a classical photon. This interpretation is only allowed with these concepts of extended causality and of classical photon; it is not possible with the concept of a continuous wave.
Both are retarded solutions and there is no causality violation. This is in contradistinction to the Li\`enard-Wiechert solutions of the usual wave field equation: the advanced and the retarded solutions representing spherical electromagnetic waves emitted, respectively, from the past (the retarded one) at $\tau_{ret}$ and from the future (the advanced one) at $\tau_{adv}$ by the electric charge, and propagating towards x on the lightcone. The advanced solution cannot be interpreted as an annihilation effect and besides, as it is well known, it violates causality.
\item $G_{f}$ is conformally invariant which can be expected because of the massless field and of its  $(1+1)D$ support-manifold, although it is being defined in a context of (3+1) dimensions, but what is really striking is that its conformal invariance remains even for $\Delta\tau\ne0$, that is for massive fields! The proof has been transferred to the Appendix D.
\item With $$P:f\rightarrow{\bar f};\qquad P:\tau+f.x\rightarrow\tau+f.x;$$
$$T:f\rightarrow-{\bar f};\qquad T:\tau+f.x\rightarrow\tau+f.x;$$
$$C:f\rightarrow-f ;\qquad C:\tau+f.x\rightarrow-(\tau+f.x),$$ where P, T, and C are respectively the parity (${\vec x}\rightarrow-{\vec x}$), the time reversal ($t\rightarrow-t$), and the charge conjugation ($\tau\rightarrow-\tau$), we can say that $G_{f}$ is invariant under the P and the T transformation, but that the C transformation changes the b ($b=\pm1$) into a -b, interchanging the solutions or the particle creation and annihilation processes. See the Figure 8.
$$P:G({\vec x},t,\tau)_{f}\rightarrow G(-{\vec x},t,\tau)_{-{\bar f}}=G({\vec x},t,\tau)_{f}$$
$$T:G({\vec x},t,\tau)_{f}\rightarrow G({\vec x},-t,\tau)_{{\bar f}}=G({\vec x},t,\tau)_{f}$$
$$C:G({\vec x},t,\tau)_{f}\rightarrow G({\vec x},t,-\tau)_{-f}$$

\hspace{-3cm}
\vglue-3cm

\hspace{-1cm}
\parbox[t]{5.0cm}{
\begin{figure}
\vglue-3cm
\hglue-1cm
\epsfxsize=400pt
\epsfbox{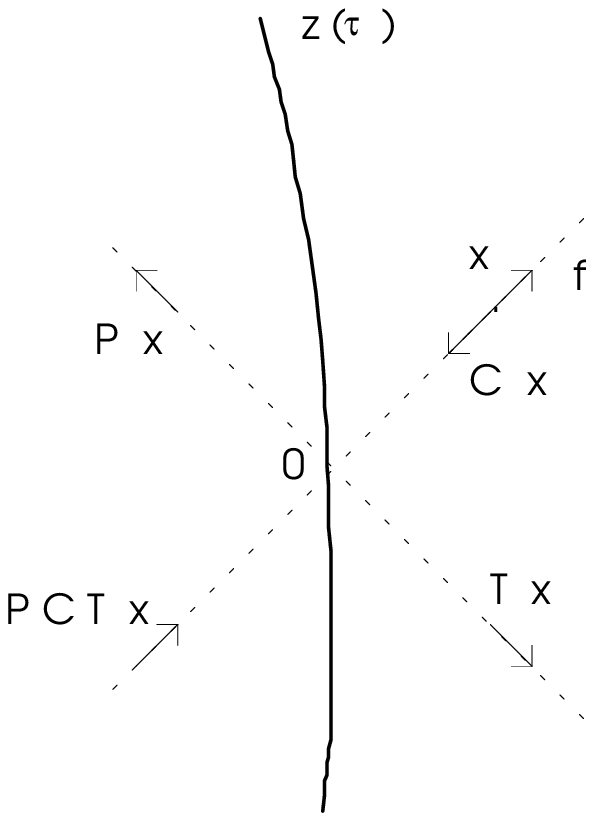}
\vglue-5cm
\end{figure}}\hfill
\\ \mbox{}
\hspace{5cm}
\parbox[t]{8.0cm}{\vglue-5cm 
Fig. 8. The $G_{f}$ symmetries under parity (P), time reversal (T), charge conjugation (C), and PCT transformations.}\\ \mbox{}
\vglue-2cm

\item One can see from (\ref{pr9}), that for $\tau=0,$ $G_{f}$ is an even function of its parameter f,
\be
\label{fef}
G_{-f}(x,\tau)=G_{f}(x,\tau).
\ee
This is an important property for retrieving the standard formalism.
\item  $G_{f}(x)$ does not depend on $x_{\hbox{\tiny T}}$, the transversal coordinate, with respect to ${\vec f}:$ $\frac{\partial}{\partial x_{\hbox{\tiny T}}}G_{f}(x)=0.$ So, there is in $G_{f}(x)$ an implicitly assumed factor $\delta^{2}(x{\hbox{\tiny T}}).$ It is a good exercise to verify that (\ref{pr9}) is a solution to (\ref{gfe}) but we leave it to the Appendix C. 
\end{itemize}

\section{\bf {RETRIEVING THE STANDARD FORMALISM}}

 The standard Li\`enard-Wiechert propagator is retrieved from (\ref{s}) and (\ref{sgf}):
\be 
\l{gg}
G(x,\tau)=\frac{1}{4\pi}\int d^{2}\Omega_{f}G(x,\tau)_{f}.
\ee
With $\tau=0$ in (\ref{pr9}) and writing $f.x=f_{4}(r\cos\theta_{f}-\varepsilon t),$ where the angle $\theta_{f}$ is defined by ${\vec f}.{\vec x}=r|{\vec f}|\cos \theta_{f},\;$ for a fixed ${\vec x}$, which we take as our coordinate z-axis, we have
\be
\l{gg2}
G(x,\tau)=\frac{2}{4\pi}\int d\phi_{f}\sin\theta_{f} d\theta\Theta(at)\delta(r\cos_{f}\theta-\varepsilon t)=\frac{1}{r}\Theta(at)\Theta(r-\varepsilon t)=\frac{1}{r}\Theta(at)\delta(r-\varepsilon t).
\ee
The factor 2 in front the integral sign is to account for the double cone ($f^{4}>0$ and $f^{4}<0,$ the past and the future cones); we normalized $|{\vec f}|$ to 1. In the last step we made use of the constraint (\ref{A}), which implies that r cannot be larger than $\epsilon t$ and so $\theta(r-\epsilon t)$ is actually a $\delta(r-\epsilon t).$
This result (\ref{gg2}) puts in evidence that the singularity $r=0$ in the standard propagator is a consequence of its average character; it is just a signal of the cone-vertex singularity. There is no physical quantity that blows to infinity as r tends to zero, according to (\ref{pr9}).

The symmetry (\ref{fef}) of $G_{f}$ makes of $A_{f}$ an even function of f. This same averaging process of (\ref{s}) reduces the f-wave equation (\ref{wef}) to the standard wave equation:
$$\frac{1}{4\pi}\int d^{2}\Omega_{f}\Box_f\;A_{f}(x,\tau)=\frac{1}{4\pi}\int d^{2}\Omega_{f}(\partial^{\mu}\partial_{\mu}-2f^{\mu}\partial_{\mu}\partial_{\tau})\;A_{f}(x,\tau)=\frac{1}{4\pi}\int d^{2}\Omega_{f}\partial^{\mu}\partial_{\mu}\;A_{f}(x,\tau),$$
as $\int d^{2}\Omega_{f}f^{\mu}\partial_{\mu}\partial_{\tau}\;A_{f}(x,\tau)=0$ by the symmetry (\ref{fef}) and so,
\be
\label{ss}
\frac{1}{4\pi}\int d^{2}\Omega_{f} \Box
_f\;A_{f}(x,\tau)=\Box\;A(x,\tau).
\ee

\section{A NEW KIND OF FIELD THEORY. CONCLUSIONS.}

The main characteristic of the extended causality is that it reduces, in a manifestly covariant way, the degrees of freedom of the (3+1)-dimensional spacetime for describing the motion of a point object  to the one degree of freedom of a (1+1)-spacetime, the straight line tangent to $f^{\mu}.$\\ At this point we have the right of asking what kind of field theory we are producing with the adoption of such a Classical-Mechanics style of causality implementation.  What are the consequences? What is the price we have to pay for having precisely what? In this section, trying to give a comprehensive perspective we may mention some results that rigorously we are going to show or prove only in still-to-come communications; we will try to make them as few as possible but for the sake of comprehensiveness some of them may be unavoidable.\\ This formalism deals only with  localized point-like fields, although these points may be part of a larger extended object. With respect to the argument that a point cannot represent a charged physical object because of the infinities associated with its self-field (think of a point electron, for example) it must be added that this extended causality cures all these problems of infinities with a point source. The interested reader is addressed to the reference \cite{hep-th/9610028}.

Another implicit characteristic of this formalism is that besides point-like fields we are dealing with point-like interactions too. They occur at points, events, defined  by the emission or absorption of other point-like fields. The interaction Lagrangian is a discrete sum of products of fields on these points.\\
The fiber $f$ is  collinear to $dx$, and being $f^{\mu}$ a constant four-vector it means that we are dealing with a free point object propagating on a straight line.

\hspace{-5cm}
\parbox[]{5.0cm}{
\begin{figure}
\epsfxsize=400pt

\epsfbox{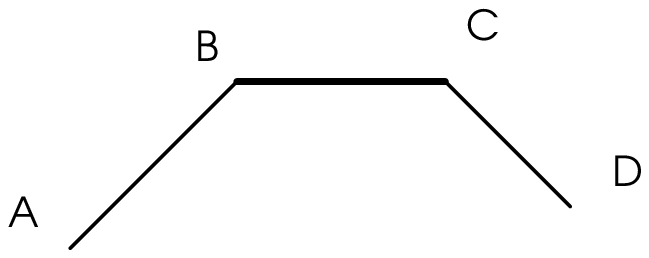}
\vglue-5cm
\end{figure}
\vglue-7cm}\\ \hfill
\mbox{}
\hspace{7cm}
\parbox[]{8.0cm}{
\vglue-11cm
Fig. 9. A polygonal trajectory of an electron. A, B, C and D are points of interactions (with an external charge) with the emission or absorption of photons. In between these points the electron travels as a free particle. Closed  cyclic trajectories are the equivalent to the stationary states of quantum mechanics. }\\
\mbox{}\\

\vglue-5cm

 This actually does not imply on any limitation at all as long as all interactions, gravitation included, are discrete in time and in space any object propagates freely between two consecutive interaction events. Its trajectory forms the a polygonal line whose vertices denote the points of interaction. See Figure 9. For a fixed point charge the space and time average of these interactions reproduces its static field. A closed cyclic trajectory is the equivalent to a stationary state in quantum mechanics. The field emitted by a charge, for example, is a point physical object that detaches from its source and moves away from it along a straight line, like a classical free particle. See the Figure 3. There is no bound (or attached to a charge) field; there is only the radiation field. There is no sense on talking about a field from an hypothetical single non-interacting charge. The static field between two charges is the spacetime average of their discrete interaction and it exists only on the straight line between the two interacting charges; they do obey to the Coulomb's law although not to the Gauss'law. See the reference \cite{hep-th/9610145} for an anticipated qualitative discussion on this. It makes no sense, therefore, to associate in any way the electron mass, or even just part of it, to its self-field. There is no connection between them. The use of differential equations for describing these discrete interactions remains valid as a useful approximation similar to the one that justifies Statistical Mechanics in the kinetic theory of gases, for example.

We must remember that $f$ is the tangent vector of a generator of a lightcone. It is a constant light-like four-vector. So, the physical picture that we are developing here is that of a discrete, in time and in space, interaction between massive point charges, mediated by the exchange of point-like massless fields propagating at a constant velocity, with the speed of light, on a straight line between a pair of interacting charges.  These are the ``classical quanta", an apparently contradictory neologism that we have to introduce for defining a classical description for the quanta. This represents a drastic change in our standard view of a classical field theory. There is no infinity and the standard formalism is regained through the equations (\ref{s},\ref{gg} and \ref{ss}).\\ 
As we are developing a discrete classical field theory with an eye on a posterior quantization someone could argue that this extended causality would not be compatible with Quantum Mechanics and the Heisenberg Principle, but actually this would be just a matter of re-interpretations. The f-formalism is remarkably closer to a relativistic quantum field formalism than it may seem at a first sight. The  field propagators $G_{f}$  have a natural interpretation in  terms of particle creation and annihilation, they can be expressed in terms of nom-commuting creation and annihilation operators, and they form a vector space: each cone generator defines a state vector; they obey to the Linear Superposition Principle.\\
We have not yet considered any actual observation or measurement of an $A_{f}$ field, but being point-like objects propagating on a cone generator they can be determined only up to uncertainties defined by the windows of our observation instrument. The necessity of a probabilistic treatment of measurements and observations is implicit in the formalism.\\ One thing is for sure: a theory based on the propagator (\ref{pr9}) excludes self interactions and vacuum fluctuation. Again, against the idea that radiative corrections are not only necessary but also confirmed by a solid experimental data one must remember that a finite theory that hopefully could replace QED must be based on equations that are modifications of the QED equations, and whose solutions, closed and finite, are equivalent to the summation of all QED radiative corrections. One may hope that these modifications on the QED equations are the one implicit in $\partial\rightarrow\nabla_{f}$ and $A\rightarrow A_{f}$ with a similar transformation for the Dirac electron spinor $\Psi\rightarrow\Psi_{f}.$ In contradistinction to what happens in QED, Figure 3 which is very similar to a Feynmann diagram does not represent just an approximation, a part of a perturbative solution; here it represents an actual elementary interaction process.

\section{\bf APPENDIX A: A GEOMETRIC PICTURE.}

The constraint (\ref{A}) represents a four-dimension hypersurface immersed in a flat five-dimension manifold, with $\Delta x^{5}=\Delta \tau.$ The $\Delta x^{5}$ of a physical object is its aging measured on its rest-frame. See the Figure 1A. For a better understanding of the meaning of $\Delta x^{5}$ as an ageing we suggest the discussion on the twin paradox in the reference \cite{hep-th/9505169}.

\begin{minipage}[t]{5.0cm}
\parbox[]{5.0cm}{
\begin{figure}
\vglue-5cm
\epsfxsize=400pt
\epsfbox{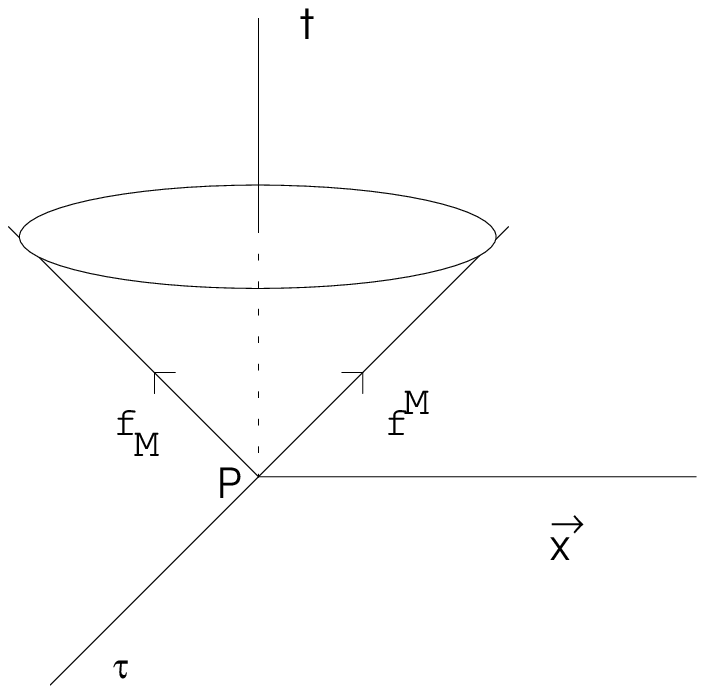}
\vglue-5cm
\end{figure}}
\end{minipage}\hfill
\hfill
\begin{minipage}[]{7.5cm}
\parbox[t]{7.5cm}{\hspace{7.0cm}
Fig. 1A. The allowed spacetime as a four-dimension hypercone embedded in a flat five-dimension manifold. $x^{5}$ is a universal invariant time: $\Delta x^{5}=\Delta\tau.$ }
\end{minipage}
\vglue-2cm
The lightcone is just a projection of this hypercone on the ($\Delta x^{5}=0)$-hyperplane. The constraint (\ref{dlc}) may be written as
\be
\label{fmd}
f_{M}dx^{M}=0,
\ee
with $M=1$ to 5, and with $M=4,5$ denoting timelike components. $\;\eta_{MN}=diag(1,1,1,-1,-1)\;$ is the metric tensor. $\;f^{M}=\frac{\Delta x^{M}}{\Delta\tau},$ and so $f^{5}=1.$$\;f^{M}$ and $f_{M}$ are, respectively, always tangent and normal to this hypercone. Therefore, (\ref{fmd}) defines a hyperplane normal to $f_{M}$ and  (\ref{fmd}) together with (\ref{A}) define a hypercone generator, tangent  to $f^{M}$. At the vertex P, a hyperplane tangent to the hypercone generator $f^{M}=({\vec f},f^{4},f^{5})$ is also orthogonal  to the hypercone generator ${\bar f}^{M}=(-{\vec f},f^{4},f^{5});$ one is the specular image of the other with respect to the plane $\Delta x^{4}=0,\;\Delta x^{5}=0,$ passing by P. Therefore,
\be
\label{fbf}
{\bar f}^{M}=-f_{M}\hspace{0.5cm}\hbox{and}\hspace{0.5cm}f^{M}=-{\bar f}_{M}.
\ee
All physical objects, regardless their masses, are constrained to remain on this hypercone, which represents the physical (accessible) four-dimension spacetime from the vertex P; the interior and the exterior of a causality cone are both prohibited regions for a physical object at the cone vertex. They are not physical spacetime.  The causality cone is, in this sense, a light-cone generalization for massive and massless fields.

\section{\bf APPENDIX B: THE GREEN'S FUNCTION.}

Using the Fourier plane-wave expansion we have
\be
\label{G}
G_{f}(x,\tau)=\int d^{5}p e^{ip_{M}x^{M}}{\tilde G}(p),
\ee
which with (\ref{delta5}) and $f^{2}=0$ in (\ref{wef}) produces
\be
\label{Gt}
{\tilde G}(p)=-\frac{1}{(2\pi)^{5}}\;\frac{1}{p^{2}-2p.f\;p_{5}},
\ee
Back with this result to (\ref{G}),
\be
G_{f}(x,\tau_{x})=-\frac{1}{(2\pi)^{5}}\int d^{5}p e^{ip_{M}x^{M}}\frac{1}{p^{2}-2p.f\;p_{5}},
\ee
and making explicit the integration on the fifth coordinate, we have
\be
G_{f}(x,\tau_{x})=\lim_{\varepsilon\to0}\frac{1}{(2\pi)^{5}}\int d^{4}p\frac{e^{ip_{\mu}x^{\mu}}}{2p_{f}} \int dp_{5}\frac{e^{ip_{5}x^{5}}}{-(p_{5}-\frac{p^{2}}{2p.f}\pm i\varepsilon)},
\ee
so that
\be
\label{pr5}
G_{f}(x,\tau_{x})=-ib\theta(bx^{5})\frac{1}{(2\pi)^{4}}\int d^{4}p\frac{e^{i(p_{\mu}x^{\mu}+\frac{p^{2}}{2p.f}})}{2p.f}.
\ee
In this equation b stands for $\pm1$, a sign that comes from the choice of the contour in a Cauchy integral, i.e. the sign of $\pm i\varepsilon$. 
Now we try to repeat the same procedure with the integration on the variable $p_{4}$ but we observe that the integrand of (\ref{pr5}) has a singularity at $p.f=0$ and that the exponent in the integrand is also only defined at this point if
\be
\label{p2f}
p^{2}{\Big |}_{p.f=0}=0,
\ee
This implies on a system of two simultaneous equations
\be
\label{p2}
p^{2}=p_{\hbox{\tiny T}}^{2}+p_{\hbox{\tiny L}}^{2}-p_{4}^{2}=0,
\ee
and 
\be
\label{pf}
p.f=p_{\hbox{\tiny L}}|{\vec f}|-p_{4}f_{4}=0,
\ee
where the subindices {\tiny L} and {\tiny T} stand, respectively, for
longitudinal and
transversal with respect to the space part $\stackrel{\rightarrow}{f}$
of f.\\
As $f^{2}=0$ we may write $\epsilon=\frac{|{\vec f}|}{f_{4}}=\pm1$, so that (\ref{pf}) becomes $p_{4}=\epsilon p_{\hbox{\tiny L}}.$ Equation (\ref{p2}) is, consequently, equivalent to
\be
\label{pt}
p_{\hbox{\tiny T}}=0.
\ee
The conditions (\ref{p2}) and (\ref{pt}) are full of physical significance: the first one requires a massless field and the second one implies that $\Delta x_{\hbox{\tiny T}}=0$; only the $x_{\hbox{\tiny L}},$ that is the longitudinal coordinate, participates in the system evolution. The field $A_{f}$ only propagates along the  fiber f. \\
For simplicity we have restricted our treatment to the case of a massless field. The general case including both massive and massless field has been discussed elsewhere \cite{cpmf}.  A consistent treatment of a massive field requires more than just replacing the wave equation (\ref{we}) by $(\Box_{f}-m^{2})A_{f}=J$ but produces the same final results \cite{cpmf}.\\ 
The equation (\ref{pr5}), as a consequence of (\ref{pt}),  is reduced to 
\be
\label{pr6}
G_{f}(x,\tau)=ib\theta(bx^{5})\frac{1}{(2\pi)^{2}}\int \frac{dp_{\hbox{\tiny L}}dp_{4}}{2f_{4}(p_{4}-\epsilon p_{\hbox{\tiny L}})}e^{i(p_{\hbox{\tiny L}}x_{\hbox{\tiny L}}+p_{4}t+\frac{p_{4}+\epsilon p_{\hbox{\tiny L}}}{2f_{4}}\tau},
\ee
with $x^{4}=t$, $x^{5}=\tau$ and after having $\frac{p^{2}}{2p.f}$ re-written as 
\be
\frac{ p_{\hbox{\tiny L}}^{2}-(p_{4})^{2}}{2p.f}=\frac{(\epsilon p_{\hbox{\tiny L}}+p_{4})(\epsilon p_{\hbox{\tiny L}}-p_{4})}{2(\epsilon p_{\hbox{\tiny L}}-p_{4})f_{4}}=\frac{\epsilon p_{\hbox{\tiny L}}+p_{4}}{2f_{4}}.
\ee
We now make explicit the integration on the coordinate $p_{4}$,
\be
\label{pr7}
G_{f}(x,\tau)=\lim_{\varepsilon\to0}ib\theta(bx^{5})\frac{1}{(2\pi)^{2}}\int dp_{\hbox{\tiny L}}dp_{4}\frac{e^{i(p_{\hbox{\tiny L}}x_{\hbox{\tiny L}}+p_{4}t+\frac{p_{4}+\epsilon p_{\hbox{\tiny L}}}{2f_{4}}\tau)}}{2f_{4}(p_{4}-\epsilon p_{\hbox{\tiny L}}\pm i\varepsilon)},
\ee
which produces
\be
\label{pr8}
G(x,\tau)_{f}=-\frac{ab}{2\pi}\theta[a(t+\frac{\tau}{2f_{4}})]\theta(bx^{5})\int \frac{dp_{\hbox{\tiny L}}}{2f_{4}}e^{ip_{\hbox{\tiny L}}(x_{\hbox{\tiny L}}+\epsilon t+\frac{\epsilon}{f_{4}} \tau)},
\ee
where $a=\pm1$ is also connected to the choice of a sign in $\pm\varepsilon.$
On the other hand
\be
\label{d}
\frac{1}{2\pi}\int \frac{dp_{\hbox{\tiny L}}}{f_{4}}e^{ip_{\hbox{\tiny L}}(x_{\hbox{\tiny L}}+\epsilon t+\frac{\epsilon \tau} {f_{4}})}=\frac{1}{2\pi\epsilon}\int \frac{\epsilon dp_{\hbox{\tiny L}}}{f_{4}}e^{i\frac{\epsilon p_{\hbox{\tiny L}}}{f_{4}}(f_{\hbox{\tiny L}}x_{\hbox{\tiny L}}+f_{4} t+\tau)}=\frac{1}{\epsilon}\delta(\tau+f.x),
\ee
and therefore, we have for (\ref{pr8})
\be
\label{pr8'}
G(x,\tau)_{f}=-\frac{ab\epsilon}{2}\theta[a(t+\frac{\tau}{2f_{4}})]\theta(b\tau)\delta(\tau+f.x).
\ee
The signs $a=\frac{\Delta t}{|\Delta t|},\;\;b=\frac{\Delta \tau}{|\Delta \tau|}\;$ and $\epsilon=-\frac{|f^{4}|}{f^{4}}$ are constrained by \cite{cpmf}
\be
\l{abe} 
ab\epsilon=-1,
\ee
which can schematically be seen from
\be
ab\epsilon=\frac{\Delta t}{|\Delta t|}\frac{\Delta \tau}{|\Delta \tau|}\frac{|f_{4}|}{f_{4}}=-\frac{\Delta t}{|\Delta t|}\frac{\Delta \tau}{|\Delta \tau|}\frac{|f^{4}|}{f^{4}}=-\frac{|\Delta \tau||f^{4}|}{|\Delta t|}=-1.
\ee
So, we can use (\ref{abe}) and the delta-function's argument for re-writing (\ref{pr8'}) as
\be
\label{pr10}
G_{f}(x,\tau_{x})=\frac{1}{2}\theta(b{\bar f}.x)\theta(b\tau)\delta(\tau+f.x),
\ee
which explicitly shows the invariance of the arguments of the theta functions.  $\quad f^{\mu}=({\vec f}, f^{4})$, ${\bar f}^{\mu}=(-{\vec f}, f^{4}).$ 
But $\theta[a(t+\frac{\tau}{2f_{4}})]=\theta(at),$ a consequence of (\ref{A}), and so $G_{f}$ can also be put as
\be
\label{pr8a}
G(x,\tau)_{f}=\frac{1}{2}\theta(at)\theta(b\tau)\delta(\tau+f.x).
\ee

\section{\bf APPENDIX C: $\partial^{\mu}\partial_{\mu}G_{f}(x)$.}

It is a good mathematical exercise to verify that (\ref{pr8a}) satisfies (\ref{gfe}). Actually, after (\ref{pt}), one can consider that there is an implicit $\delta^{2}(x_{\hbox{\tiny T}})$ on the right-hand side of (\ref{pr8a}), or equivalently that (\ref{gfe}) is replaced by
\be
\label{gfe'}
\eta^{\mu\nu}\nabla_{\mu}\nabla_{\nu}G_{f}(x-y,\tau_{x}-\tau_{y})=\delta(t_{x}-t_{y})\delta(\tau_{x}-\tau_{y})\delta(x-y)_{\hbox{\tiny L}}.
\ee
This checking is made almost trivial if we make a judicious  handling of  constraints:
\be
\partial_{\mu}G_{f}(x)_{f}=\frac{1}{2}\partial_{\mu}\{\theta(at)\theta(b\tau)\delta(\tau+f.x)\}=\frac{1}{2}(\partial_{\mu}-f_{\mu}\partial_{\tau}){\Big \{}\theta(at)\theta(b\tau){\Big \}}{\Big |}_{\tau+f.x=0}=
\ee
\be
=\frac{1}{2}{\Big \{}-af_{\mu}\delta(\tau)\theta(b\tau)+b\delta_{\mu4}\delta(t)\theta(\tau){\Big \}}{\Big |}_{\tau+f.x=0}.
\ee
\be
\nonumber
\partial^{2}_{\nu\mu}G_{f}(x)=\frac{1}{2}(\partial_{\nu}-f_{\nu}\partial_{\tau}){\Big \{}-af_{\mu}\theta(b\tau){\Big |}_{\tau=0}+b\delta_{\mu4}\theta(\tau){\Big |}_{t=0}{\Big \}}{\Big |}_{\tau+f.x=0}=
\ee
\be
\partial^{2}_{\nu\mu}G_{f}(x)=\frac{1}{2}{\Big \{}-af_{\mu}b\delta_{\nu4}\delta(\tau){\Big |}_{t=0}-b\delta_{\mu4}af_{\nu}\delta(\tau){\Big |}_{t=0}{\Big \}}{\Big |}_{\tau+f.x=0}.
\ee
So,
\be
\partial^{\mu}\partial_{\mu}G_{f}(x)=-abf_{4}\delta(t)\delta(\tau)\delta(\tau+f.x)=|f_{4}|\delta(t)\delta(\tau)\delta(|{\vec f}|x_{\hbox{\tiny L}})=\delta(t)\delta(\tau)\delta(x_{\hbox{\tiny L}}).
\ee

\section{\bf APPENDIX D: CONFORMAL INVARIANCE.}

Let us consider the coordinate transformation
\be
\l{ct}
x^{\mu}\rightarrow y^{\mu}=\frac{ax^{\mu}+b^{\mu}}{(ax+b)^{2}}
\ee
which for simplicity, can with a change of variables be reduced to just
\be
\l{ctr}
x^{\mu}\rightarrow y^{\mu}=\frac{x^{\mu}}{x^{2}}.
\ee
It implies on 
\be
\l{xy}
x^{2} y^{2}=1,
\ee
and to
\be
\l{tt}
\tau_{x}\tau_{y}=\pm1
\ee
with $(\tau_{x})^{2}=x^{2}$ and $(\tau_{y})^{2}=y^{2}.$ 
So,
\be
\l{dx}
dx^{\mu}=\frac{\partial x^{\mu}}{\partial y^{\nu}}=X^{\mu}_{\nu}dy^{\nu},
\ee
and 
\be
\l{dy}
dy^{\mu}=\frac{\partial y^{\mu}}{\partial x^{\nu}}=Y^{\mu}_{\nu}dx^{\nu},
\ee
with
\be
\l{Y}
Y^{\mu}_{\nu}= y^{2}\delta^{\mu}_{\nu}-2y^{\mu}y_{\nu}=\frac{X^{\mu}_{\nu}}{(x^{2})^{2}};
\ee
\be
\l{X}
X^{\mu}_{\nu}= x^{2}\delta^{\mu}_{\nu}-2x^{\mu}x_{\nu}=\frac{Y^{\mu}_{\nu}}{(y^{2})^{2}};
\ee
\be
\l{XY}
X^{\mu}_{\alpha}Y^{\alpha}_{\nu}= \delta^{\mu}_{\nu};
\ee
\be
\l{dtau}
d\tau_{x}=\frac{d\tau_{y}}{(\tau_{y})^{2}}.
\ee
\be
\l{f}
f_{\mu}=-\frac{\partial\tau_{x}}{\partial x^{\mu}}=-\frac{\partial\tau_{x}}{\partial\tau_{y}}\frac{\partial\tau_{y}}{\partial y^{\nu}}\frac{\partial y^{\nu}}{\partial x^{\mu}}=-\frac{Y^{\nu}_{\mu}}{(\tau_{y})^{2}}\;g_{\nu},
\ee
with 
\be
\l{g}
g_{\nu}=-\frac{\partial\tau_{y}}{\partial y^{\nu}},
\ee
or using (\ref{tt}) and (\ref{XY}) for writing
\be
\l{gf}
g_{\alpha}=-\frac{X^{\nu}_{\alpha}}{(\tau_{x})^{2}}f_{\nu}.
\ee
So,
\be
\l{dl}
d\tau_{x}+f.dx=-\frac{1}{(\tau_{y})^{2}}(d\tau_{y}+g.dy).
\ee

\end{document}